\documentclass{article}
\usepackage{caption}
\usepackage{subcaption}
\usepackage{cite}
\usepackage{algorithmic}
\usepackage[sort,numbers]{natbib}

   \usepackage{graphicx}
 
\usepackage[cmex10]{amsmath}
\usepackage{amsfonts}
\usepackage{array}

\newcommand{\beq}{\begin{equation}}
\newcommand{\eeq}{\end{equation}}
\newcommand{\ind}[1]{I_{\{#1\}}}
\newcommand{\ex}{E}
\newcommand{\bfx}{\mathbf{X}}
\newcommand{\bfa}{\mathbf{a}}
\newcommand{\bfb}{\mathbf{b}}
\newcommand{\bfw}{\mathbf{W}}
\newcommand{\bfy}{\mathbf{y}}
\newcommand{\bfz}{\mathbf{Z}}
\newcommand{\bfu}{\mathbf{U}}
\newcommand{\bff}{\mathbf{f}}

\begin{document}

\title{Series Expansion Approximations of Brownian Motion
  for Non-Linear Kalman Filtering of Diffusion Processes}

\author{Simon~Lyons,
        Simo~S\"{a}rkk\"{a},
        and~Amos~Storkey% <-this % stops a space   
\thanks{S.M.J Lyons and A.J. Storkey are with
the department of Informatics, Edinburgh University. E-mail: s.lyons-4@sms.ed.ac.uk, a.storkey@ed.ac.uk}% <-this % stops a space
\thanks{Simo S\"{a}rkk\"{a} is with the Department of Biomedical Engineering and Computational Science, Aalto University. E-mail: simo.sarkka@aalto.fi }% <-this % stops a space
%\thanks{Manuscript received April 19, 2005; revised January 11, 2007.}
}

\maketitle

\begin{abstract}
%\boldmath
In this paper, we describe a novel application of sigma-point methods to continuous-discrete filtering. The nonlinear continuous-discrete filtering problem is often computationally intractable to solve. Assumed density filtering methods attempt to match statistics of the filtering distribution to some set of more tractable probability distributions. Filters such as these are usually decompose the problem into two sub-problems. The first of these is a prediction step, in which one uses the known dynamics of the signal to predict its state at time $t_{k+1}$ given observations up to time $t_k$. In the second step, one updates the prediction upon arrival of the observation at time $t_{k+1}$.

The aim of this paper is to describe a novel method that improves the prediction step. We decompose the Brownian motion driving the signal in a generalised Fourier series, which is truncated after a number of terms. This approximation to Brownian motion can be described using a relatively small number of Fourier coefficients, and allows us to compute statistics of the filtering distribution with a single application of a sigma-point method.

Assumed density filters that exist in the literature usually rely on discretisation of the signal dynamics followed by iterated application of a sigma point transform (or a limiting case thereof). Iterating the transform in this manner can lead to loss of information about the filtering distribution in highly non-linear settings. We demonstrate that our method is better equipped to cope with such problems.
\end{abstract}

%\begin{IEEEkeywords}
%Nonlinear Filtering, Sigma point, unscented transform, Kalman filter,
%\end{IEEEkeywords}

% \begin{center} \bfseries EDICS Category: SSP-FILT MDS-ALGO MDS-FILT NSP-NLIN \end{center}
% \fi

\section{Introduction}
Stochastic differential equations (SDEs) provide a natural way to describe the evolution of systems that are inherently noisy, or contain unknown phenomena that can be modelled as stochastic processes \cite{Jazwinski1970,Oksendal2007}. Suppose that the evolution of an idealised system could be modelled with the ordinary differential equation (ODE)

\beq
\label{ode}
\frac{d \bf\bfx_t}{dt} = \bfa(\bfx_t),
\eeq
where $\bfx_t \in \mathbb{R}^n$ is the state of the system, and $\mathbf{a}:\mathbb{R}^n \rightarrow \mathbb{R}^n$. 
Roughly speaking, to construct an SDE, one adds a `white' driving noise to the dynamics of an ODE. From the modelling perspective, the purpose of the noise is to capture deviations from the ideal deterministic model. The amplitude of this driving noise may potentially depend on the current state $\bfx_t$ of the system. The result is a differential equation

\beq
\label{naivesde}
\frac{d\bfx_t}{dt} = \bfa(\bfx_t) + \bfb(\bfx_t)\dot{\bfw}_t,
\eeq
where $\dot{\bfw}_t \in \mathbb{R}^d$ is Gaussian white noise, and $\mathbf{b}:\mathbb{R}^n \rightarrow \mathbb{R}^{n\times d}$. Because of the highly irregular nature of continuous-time white noise, one needs to be careful when defining this equation mathematically. In order to do this, the usual approach is re-write \eqref{naivesde} as an integral equation and interpret the second term on the right as an It\^o stochastic integral \cite{Karatzas1991,Oksendal2007}:
\beq
\label{integraleqn}
\bfx_t = \bfx_0 + \int_0^t \bfa(\bfx_u)du + \int_0^t \bfb(\bfx_u)d\bfw_u.
\eeq

This allows us to interpret the dynamics as an It\^o \emph{stochastic differential equation}:
\begin{align}
\label{SDE}
d\bfx_t  &=  \bfa(\bfx_t)dt + \bfb(\bfx_t) d\bfw_t, \qquad \bfx_0 = \mathbf{x}_0.
\end{align}
The solution $\bfx_t$ will then be an \emph{It\^o diffusion process}.
Here, the term $d\bfw_t$ denotes the infinitesimal change in a $d$-dimensional Brownian motion. We assume $\bfw$ is a standard Brownian motion, so that its components are independent with variance $t$ at time $t$ (we use the convention that a vector-valued stochastic process is represented with an upper-case bold letter, whereas a stochastic process \emph{evaluated at a given time} also has a subscript). In situations where we need to refer to, say, the $k$-th component of the vector $\bfx_t$, we use the notation $\bfx_k(t)$.

One must make some assumptions about $\bfa$ and $\bfb$ to ensure Equation \eqref{SDE} has a unique solution. If both functions are globally Lipschitz and grow at most linearly, one is assured that this will be the case \cite{Karatzas1991}.

It is often the case that one cannot observe the process $\bfx$ directly---instead, one must rely on discrete-time, noisy observations $\{\mathbf{Y}_{t_k} \in \mathbb{R}^s\}_{k \geq 1}$ of the process. In mathematical terms, the model for measurements of this type can often be written as
\beq
\mathbf{Y}_{t_k} = \mathbf{h}(\bfx_{t_k}) + \mathbf{V}_{t_k},
\eeq
for some known `observation function' $\mathbf{h}$ with Gaussian measurement noise $\mathbf{V}_{t_k} \sim \mathcal{N}(0,\mathbf{R}_k)$. One is then often faced with the task of computing the expectation $\ex[\phi(\bfx_{t}) | \mathbf{Y}_{t_1}, \dots, \mathbf{Y}_{t_f}]$, where $t \ge t_f$ for some given function $\phi$. This is known as the \emph{continuous-discrete filtering problem}. For simplicity, we assume that the conditional distribution of $\bfx_t$ has a density with respect to the Lebesgue measure. For filtering problems where this is not the case, such as when part of the system is observed without error, much of our analysis can be applied with only minor modifications. The estimation problem can be solved for arbitrary $\phi$ provided that we can compute the filtering density $p_{\bfx_t}\left( x \ | \{\mathbf{Y}_{t_k} : {t_k \leq t} \} \right)$ for all $t$. This latter approach is often called the probabilistic or Bayesian approach to the filtering problem \cite{Jazwinski1970}.

It is only in a small number of special cases that the conditional distribution of $\bfx_t$ can be described using a finite number of parameters. When the SDE is linear and the function $\mathbf{h}$ in the measurement model is linear, then the Kalman filter can be used to compute the exact solution \cite{Kalman1960}. Certain other filtering problems also admit closed-form solutions (see, for example, the Bene\u{s} filter \cite{Baine2009}). However, closed-form filters are rare, and in most cases one must approximate the filtering distribution in some manner. For example, one can discretise the signal and employ a particle filter \cite{Sarkka2008,Fearnhead2008,Murray2011,Kotecha2003}, which uses Monte Carlo samples to approximate the filtering distribution. Other approaches include variational filtering \cite{Friston2008a}, homotopy filtering \cite{Daum2008}, and path integral filtering \cite{Balaji2009}.

Another general technique is to take a parametric set of tractable densities (for example a set of densities within the exponential family) and find the density within that set that most closely matches the filtering density. This approach, introduced in \cite{Kushner1967}, is known as \emph{assumed density filtering}.

In this paper, we will attempt to compute statistics of the Gaussian distribution that most closely matches the filtering distribution. This particular special case of assumed density filtering is known as \emph{Gaussian filtering} \cite{Ito2000}. There are a number of ways to approach the problem. The \emph{extended Kalman filter} (EKF) \cite{Jazwinski1970} uses a Taylor series approximation to the non-linearities in SDE and measurement model. The \emph{unscented Kalman filter} (UKF) \cite{Julier1995,Julier2000,Sarkka2007} uses a set of sigma-points for computing the mean and covariance of the Gaussian approximation. Quadrature and cubature based filters \cite{Ito2000,Arasaratnam2009,Arasaratnam2010,Saerkkae2013} use Gaussian numerical integration for computing the mean and covariance. The Gaussian assumption is a natural one when the filtering distribution is known to be unimodal. However, it may lead to significant errors for certain multimodal distributions. It is not advisable to apply a Gaussian filter blindly, without considering the possibility of encountering a multimodal filtering distribution.

The commonly used approaches to filtering in continuous-discrete systems can be divided into two categories: one possibility is that the SDE is first discretised using methods such as It\^o--Taylor series or a stochastic Runge--Kutta discretisation \cite{Kloeden1999}, \cite{Murray2011}. Discrete-time filtering algorithms are the applied to the discretised process. The alternative is that an approximate filter is formed that operates in continuous time, and that filter is discretised. The relative merits of these approaches were recently studied in \cite{Sarkka}.

In this paper, we take a different approach. We begin by fixing an interval $[0,T]$, which will typically be the time between observations. Observe that one can view the Brownian motion $\bfw$  as a random element of the Hilbert space $L^2[0,T]$. It is an inherently infinite-dimensional object. However, one can construct a finite-dimensional approximation of $\bfw$ by projecting it onto a finite-dimensional subspace of $L^2[0,T]$ \cite{Stroock2010}. We use the projection as the driving noise in an approximation of the original signal. The transition map is then approximated as a function that satisfies a certain ordinary differential equation. We refer to the new filter as the \emph{series expansion unscented Kalman filter} (SE-UKF).

Ideas of this type were first explored by Wong and Zakai \cite{Wong1965}. Similar ideas have been explored in \cite{Corlay2010}, \cite{Corlay2011} in the context of variance reduction for Monte-Carlo simulation, and in \cite{Lyons2012} in the context of parameter estimation.
In this framework, one can interpret the approximation we use as the image of an $N \times d$-dimensional standard normal distribution under a nonlinear transform. This suggests the possibility of using sigma-point methods such as the unscented transform to construct a Gaussian filter.

Gaussian filters that currently exist in the literature typically rely on discretisation of the signal. The time-$t$ distribution of the discretised signal is repeatedly projected onto the set of Gaussian distributions, for example through moment matching or by minimising some form of generalised metric as in \cite{Archambeau2008}. Our methodology avoids repeated projection onto the space of Gaussian random variables during the prediction phase. For this reason we expect our new prediction step to outperform the prediction steps of existing methods when the inference problem is sufficiently nonlinear.

Our paper is structured as follows. In Section II, we describe our model of the filtering problem and briefly review some methods that are used in the literature at present. In Section III, we describe our method of approximating the time-$t$ marginal distribution of a diffusion process, and we show how the approximation can be exploited to construct a novel Gaussian filter. The accuracy of this approximation is investigated in Section IV, and we show that our filter performs well on a high-dimensional nonlinear problem. In Section V, we review our work and discuss some questions that arise as a result of the study.

\section{Gaussian Filtering}
\subsection{Sigma point approximations}
One widely-used approach to Gaussian filtering relies on so-called `sigma point' approximations, perhaps the best known of which is the unscented transform \cite{Julier1995,Julier2000}. Given a random variable $\bfu$ and a function $\bff$, we wish to approximate the distribution of $\bff(\bfu)$. In order to accomplish this, one chooses a number of points $\{\sigma_i\}$ that represent the distribution of $\bfu$ in some sense.

We will restrict our exposition to the case where $\bfu$ has an $n$-dimensional multivariate normal distribution, and we wish to fit a multivariate normal distribution to $\bff(\bfu)$. Suppose $\bfu$ has mean $\mathbf{m}$ and covariance $\mathbf{P}$. The unscented transform uses $2n+1$ sigma points, which are constructed as follows. One chooses two tuning parameters $\alpha$ and $\kappa$, then sets $\lambda = \alpha^2( n+\kappa) - n$. The sigma points are then defined by the following expressions:
\begin{align}
 \sigma^0 &= \mathbf{m}, \\
\sigma^i &= \mathbf{m} + (\sqrt{(n + \lambda)\mathbf{P}})_{* i}, \quad 1 \leq i \leq n, \\
\sigma^{n+i} &= \mathbf{m} - (\sqrt{(n + \lambda)\mathbf{P}})_{* i}, \quad 1 \leq i \leq n.
\end{align}

Here $(\sqrt{\mathbf{P}})_{* i}$ is the $i$-th column of the matrix square root of $\mathbf{P}$, defined as any matrix that satisfies $\mathbf{P} = \sqrt{\mathbf{P}} \sqrt{\mathbf{P}}^\top$. The sigma points are determined once one fixes a specific matrix square root (e.g. the Cholesky decomposition of $\mathbf{P}$).

The mean and covariance of $\bff(\bfu)$ are approximated by a weighted average of the sigma-point images. Define $\mathcal{Y}_i = \bff(\sigma^i)$, and set
\beq
\label{spmean}
\mu := \sum_{i=0}^{2 n} w^{(m)}_{i} \mathcal{Y}_i \simeq \ex[\bff(\bfu)].
\eeq
We can then make the approximations 
\begin{align}
\notag
\label{spvar}
& \ex[(\bff(\bfu) - \ex[\bff(\bfu)]) (\bff(\bfu) - \ex[\bff(\bfu)])^\top ] \\   
 & \simeq  \sum_{i=0}^{2 n} w^{(c)}_{i}
 \left( \mathcal{Y}_i - \mu \right)\left( \mathcal{Y}_i - \mathbf{\mu} \right)^\top, \\
\notag
 &\ex[(\bfu - m) (\bff(\bfu) - \ex[\bff(\bfu)])^\top ] \\
& \simeq    
  \sum_{i=0}^{2 n} w^{(c)}_{i}
  \left( \sigma^i - \mathbf{m} \right)\left( \mathcal{Y}_i - \mu \right)^\top.
\end{align}

The weights depend on a third tuning parameter $\beta$, and are given by
\begin{align}
\notag
 w^{(m)}_0 &= \frac{\lambda}{n+\lambda}, \\ \notag
 w^{(c)}_0 &= \frac{\lambda}{n+\lambda} + (1 - \alpha^2 + \beta),\\\notag
 w^{(m)}_i &= \frac{1}{2(n+\lambda)} \qquad i = 1, \dots, 2n,\\
 w^{(c)}_i &= \frac{1}{2(n+\lambda)}\qquad i = 1, \dots, 2n.
\end{align}
It is well known that the unscented transform matches the mean of $\bff(\bfu)$ exactly when $\bff$ is a polynomial of degree three or less. In general, errors in the estimate of the mean are introduced only by the fourth and higher terms in the Taylor expansion of $\bff$ \cite{Julier04unscentedfiltering}.

\subsection{Sigma point Kalman filters for diffusion processes}
\label{spkf}

In the Gaussian filtering paradigm, of which the unscented Kalman filter (UKF) is a special case, the filtering problem is reduced to computation of the conditional mean and covariance of the filtering distribution: 
\beq
\mathbf{m}_t = \ex \left[\bfx_t \mid \{\mathbf{Y}_{t_k} : {t_k \leq t} \} \right]
\eeq
and
\beq
\mathbf{P}_t = \text{Cov} \left[\bfx_t \mid \{\mathbf{Y}_{t_k} : {t_k \leq t} \} \right].
\eeq

This procedure is usually divided up into two steps: the \emph{prediction} step and the \emph{update} step. In the prediction step, we begin with an estimate of the mean and covariance of the filtering distribution at time $t_{k-1}$. We then use the known dynamics of $\bfx$ to compute the mean and variance of the filtering distribution the instant before the next observation arrives. 

Upon arrival of the observation at time $t_k$, we proceed to the \emph{update} step. In this step, we update our estimate of the mean and variance of the filtering distribution using information from observation $\mathbf{Y}_{t_k}$.

It is usually necessary to approximate the conditional mean and covariance: for a general nonlinear diffusion, the moments are only known in terms of the solution of a partial differential equation known as the Fokker-Planck equation \cite{Jazwinski1970,Oksendal2007}. In dimensions higher than three, the Fokker-Planck equation is typically numerically intractable.

The simplest application of the UKF to a diffusion relies on discretisation of the process $\bfx$. Suppose that at time $t_{k-1}$ we have an estimate of $\mathbf{m}_{t_{k-1}}$ and $\mathbf{P}_{t_{k-1}}$. In the prediction step, our aim is to compute an estimate of $\mathbf{m}_t$ and $\mathbf{P}_t$ the instant before the next observation arrives.
 
We divide the time interval $[t_{k-1},t_{k}]$ into a number of sub-intervals of length $\Delta t$ (for clarity, we will discuss the interval $[0,t_1]$ here). We then approximate the SDE \eqref{SDE} on the grid $\{\bfx_{\Delta t}, \bfx_{2 \Delta t}, \dots\}$ via the relation
\beq
\bfx_{(j+1)\Delta t} = \mathbf{f}(\bfx_{j \Delta t},\bfz_j),
\eeq
where $\bfz_0,\bfz_1,\ldots$ is a suitable sequence of Gaussian random variables. Here, $\mathbf{f}$ is a transition function that depends on the method of discretisation, and $\bfz_{k}$ is typically draw from a spherical Gaussian distribution of dimension $d$. For example, in the \emph{Euler-Maruyama} scheme \cite{Kloeden1999}, 
\beq
\label{euler}
\mathbf{f}(\bfx_{j \Delta t}, \bfz_j) = \bfx_{j\Delta t} + \bfa(\bfx_{j\Delta t})\Delta t + \bfb(\bfx_{j\Delta t})\sqrt{\Delta t}\bfz_{j},
\eeq
where $\bfz_{j} \sim \mathcal{N}(0,\mathbf{I}_d)$.

In this sense, $\bfx_{(j+1)\Delta t}$ is the image of $(\bfx_{j\Delta t},\bfz_j)$ under a nonlinear transform $\bff$. Given a Gaussian approximation to $\bfx_{j \Delta t}$, one can apply the unscented transform to $\bff$ to find a Gaussian approximation of $\bfx_{(j+1)\Delta t}$. One proceeds iteratively until $t_{k}$, at which point the prediction phase ends and we proceed to the update phase. Instead of the Euler--Maruyama method, one can in some circumstances use higher order It\^o--Taylor expansions, stochastic Runge--Kutta methods or various other methods \cite{Kloeden1999}.

Alternatively, one can take a limit as $\Delta t \rightarrow 0$ instead of iteratively applying the unscented transform at the prediction. By doing so, one recovers a system of differential equations for the predictive mean and covariance (see, e.g., \cite{Sarkka2007,Saerkkae2013}):
\begin{align}
\label{predictivemomentsode}
\notag
\frac{d\mathbf{m}_t^-}{dt} &= \ex[\bfa(\bfx_t^-)] \\ \notag
\frac{d\mathbf{P}_t^-}{dt} &= \ex[\bfa(\bfx_t^-) (\bfx_t^- - \mathbf{m}_t^-)^\top] \\ \notag &+ 
\ex[(\bfx_t^- - \mathbf{m}_t^-) \bfa^\top(\bfx_t^-)] \\ &+ \ex[\bfb(\bfx_t^-) \bfb^\top(\bfx_t^-)],
\end{align}
where the expectations are taken with respect to the Gaussian distribution
$\bfx_t^- \sim \mathcal{N}(\mathbf{m}^-_{t}, \mathbf{P}_{t}^-)$.

When $t = t_k$, we will make a new observation. We must then update our predictive distribution with the new information from that observation. Let $\mathbf{m}^-_{t_{k}}$ and $\mathbf{P}^-_{t_{k}}$ be the mean and covariance of the predictive distribution immediately before the new observation arrives. We form an approximation $\hat{\bfx}_{t_k}^-$ of the signal, which is Gaussian with the predictive mean and covariance. The update equations using the linear minimum mean square error estimator are as follows:
\begin{align}
\label{six}
\notag
\mu_{k} &= \ex[h(\hat{\bfx}_{t_k}^-)] \\ \notag
\mathbf{S}_{k} &= \ex[(h(\hat{\bfx}_{t_k}^-) - \mu_{k})(h(\hat{\bfx}_{t_k}^-) - \mu_{k})^\top] + \mathbf{R}_k \\ \notag
\mathbf{C}_{k} &= \ex[(\hat{\bfx}_{t_k}^- - \mathbf{m}^-_{t_{k}})(h(\hat{\bfx}_{t_k}^-) - \mu_{k})^\top] \\ \notag
\mathbf{K}_{k} &= \mathbf{C}_{k} \mathbf{S}_{k}^{-1} \\ \notag
\mathbf{m}_{t_{k}} &= \mathbf{m}_{t_k}^- + \mathbf{K}_k(\mathbf{Y}_{t_k} - \mu_k) \\
\mathbf{P}_{t_{k}} &= \mathbf{P}_{t_{k}}^- - \mathbf{K}_{k} \mathbf{S}_{k} \mathbf{K}_{k}^\top,
\end{align}
The updated distribution has mean $\mathbf{m}_{t_{k}}$ and covariance $\mathbf{P}_{t_{k}}$. When the observation function $\mathbf{h}$ is nonlinear, one can apply the unscented transform to $\mathbf{h}(\hat{\bfx}_{t_k}^-)$ to compute an approximation of $\mu_k$, $\mathbf{S}_k$, and $\mathbf{C}_k$ in \eqref{six}. More complex update rules that have been tuned for numerical stability are also known in the literature \cite{Grewal2011}.

In Section \ref{s3}, we describe a new method of approximating the predictive mean and variance by constructing a function $\bff$ so that $\bfx_t \approx \bff(t, \bfx_0, \bfz_1, \dots, \bfz_N)$, where the random variables $\{\bfz_i\}$ follow a standard normal distribution. We can apply sigma point methods to $\bff$ to estimate the mean and variance of $\bfx_t$. The image of each sigma point is computed by solving an ordinary differential equation. 
Our method requires \emph{one} application of the unscented transform per observation (though this is generalised in Section \ref{stepsize}). This is in contrast to the standard UKF, which discretises the system first, then iteratively applies the unscented transform at each timestep.

\section{Sigma Point Filtering via Smooth Approximations of Stochastic Differential Equations}
\label{s3}
\subsection{Series expansions of Brownian motion}

We now describe a method for obtaining a smooth approximation of Brownian motion by decomposing it in a generalised Fourier series. We aim to use the smooth approximation as a driving function in a differential equation. This will enable us to approximate a nonlinear stochastic differential equation with a \emph{randomised ordinary differential equation}, which will prove to be computationally tractable to work with. This approximation was used as the basis of a Markov chain Monte Carlo algorithm for Bayesian parameter estimation of a nonlinear diffusion in \cite{Lyons2012}.

Suppose $\bfw = (W^{(1)}, \dots, W^{(d)})$ is a standard $d$-dimensional Brownian motion, and let $\{\phi_i\}_{i \geq 1}$ be an orthonormal basis of $L^2([0,T], \mathbf{R})$. We use the notation $I$ to denote the indicator function. That is, $\ind{[0,t]}(u) = 1$ when $0 \leq u \leq t$, and $\ind{[0,t]}(u) = 0$ otherwise. One can construct a series expansion of $\bfw$ in terms of the basis functions $\{\phi_i\}$ as follows \cite{Luo2006}:
\begin{align}
\label{bmseries}
\notag
\bfw_t &= \int_0^T \ind{[0,t]}(u)d\bfw_u \\ \notag
&= \int_0^T \left(\sum_{i=1}^\infty \langle \ind{[0,t]}, \phi_i \rangle \phi_i(u)\right) d\bfw_u \\
&= \sum_{i=1}^\infty \left(\int_0^T \phi_i(u) d\bfw_u \right) \int_0^t \phi_i(u)du.
\end{align}
We use the standard inner product on $L^2[0,T]$, which is defined as
\beq
\langle f,g \rangle = \int_0^T f(u)g(u)du.
\eeq

For ease of notation, we set
\beq
\label{zintegral}
\bfz_i = \int_0^T \phi_i(u) d\bfw_u.
\eeq

The stochastic integrals are i.i.d $d$-dimensional standard normal. We can see this by noting that he basis functions are deterministic, and hence the integrals are Gaussian. We have
\beq
\ex\left [ \bfz_i \right] = 0,
\eeq
and, by  It\^o's isometry,
\begin{align}
\text{Cov}(\bfz_i,\bfz_j) 
= \left(\int_0^T \phi_i(u) \phi_j(u) du\right) \mathbf{I}_d = \delta_{ij}\mathbf{I}_d.
\end{align}
Here, $\mathbf{I}_d$ is the $d \times d$ identity matrix.

We conclude that
\beq
\label{dbmseries}
\bfw_t = \sum_{i=1}^\infty \bfz_i \int_0^t \phi_i(u)du.
\eeq
We can obtain an approximation of a Brownian sample path by drawing i.i.d samples $\bfz_i$ from a standard normal distribution and truncating the sum in \eqref{dbmseries}. This allows us to describe a Brownian sample path approximately in terms of a finite number of variates. This representation is crucial for our implementation of sigma-point inference methods.

\subsection{Series Expansion Approximation of SDE}
In order to approximate the diffusion $\bfx$, we truncate the series expansion \eqref{dbmseries} after $N$ terms, and use the resulting smooth process as an approximation of Brownian motion. We replace the stochastic integral in Equation \eqref{integraleqn} with the time derivative of the truncated process:
\beq
\label{sdeapprox}
\hat{\bfx}_t  = \bfx_0 + \int_0^t \bfa(\hat{\bfx}_u)du + \sum_{i=1}^N  \int_0^t \bfb(\hat{\bfx}_u) \bfz_i \phi_i(u)du.
\eeq
Since $\hat{\bfx}$ is driven by a finite linear combination of basis functions, the resulting process is differentiable. We can therefore interpret $\hat{\bfx}$ as the solution to an ordinary differential equation with a random driving function.
\beq
\label{randode}
\frac{d\hat{\bfx}_t}{dt} = \bfa(\hat{\bfx}_t) + \sum_{i=1}^N \bfb(\hat{\bfx}_t) \bfz_i \phi_i(t), \qquad \hat{\bfx}_0 = x_0.
\eeq
Approximations of this type were first investigated by Wong and Zakai \cite{Wong1965}, who showed that in the one-dimensional case, $\hat{\bfx}_t$ converges to the Stratonovich solution of the stochastic differential equation \cite{Kloeden1999}. Convergence issues are discussed in a more general multidimensional setting in the appendix. 

The approximation \eqref{randode} has the advantage of re-casting an infinite dimensional problem in finite-dimensional terms. We can view the solution of \eqref{randode} as a function
\beq
\hat{\bfx}_t = \hat{\bfx}(t, x_0, \bfz_{1} \dots, \bfz_{N}).
\eeq
In essence, the time-$t$ distribution of the process $\hat{\bfx}$ can be interpreted as the image of a $d \times N$-dimensional Gaussian distribution under a nonlinear transform. This is precisely the setting for which sigma-point methods were designed.

\subsection{The series expansion filter}

Our algorithm proceeds as follows. We assume we have a Gaussian approximation $\mathcal{N}(\mathbf{m}_{t_{k-1}}, \mathbf{P}_{t_{k-1}})$ to the filtering distribution at time $t_{k-1}$. We wish to compute the filtering distribution at time $t$. If $t<t_k$, we compute the predictive distribution. If $t = t_k$, we must also update the predictive distribution with the information gained from our observation $\mathbf{Y}_{t_k}$. 

We choose a  set $\{\sigma_j\}$ of  sigma points to represent the joint distribution of the state and the random coefficients $\{\bfz_i\}$ in \eqref{randode}. Each sigma point can be thought of as a vector of dimension $n + d\times N$,
\beq
\sigma^j = (\sigma_x^j, \sigma_z^j).
\eeq

Here, the first $n$ elements $\sigma_x^j$ of the vector $\sigma^j$ are the sigma points for the initial condition for the ODE \eqref{randode}, that is, the sigma points that represent $\mathcal{N}(\mathbf{m}_{t_{k-1}}, \mathbf{P}_{t_{k-1}})$. The remaining $d \times N$ elements $\sigma_z^j$ are the sigma points corresponding to an $N$-term expansion of a $d$-dimensional Brownian motion. Together, these data determine an initial value problem. For each sigma point $\sigma^j$, we solve the ordinary differential equation \eqref{randode}. The initial condition is $\bfx_{t_{k-1}} = \sigma_x^j$ and the coefficients representing $\{\bfz_i\}_{i \leq N}$ are formed from the appropriate subvectors of $\sigma_z^j$ (each one having length $d$). At time $T$, the solution is an $n$-dimensional vector
\beq
\hat{\bfx}^{j}_{T} = \hat{\bfx}(T, \sigma^j_x, \sigma^j_z).
\eeq 

We treat the solution at time $T$ of the initial value problem as the image of the sigma point $\sigma^j$. The set of vectors $\{\hat{\bfx}_t^j\}$ can be thought of as a discrete approximation to the predictive distribution. We can use these vectors to compute an estimate of $\mathbf{m}_t$ and $\mathbf{P}_t$, though the specific computation depends on the choice of sigma-point method. This methodology is in marked contrast to the sigma point Kalman filters of Section \ref{spkf}. These rely on discretisation of the signal dynamics and sigma point approximation of the Brownian increment $\bfw_{t + \Delta t} - \bfw_t$ at each timestep, or a limiting case of this discretisation as $\Delta t \rightarrow 0$.

We summarise our algorithm in pseudocode as follows:

\begin{algorithmic}
\FOR{$k = 1:m$}
\STATE{Set $\mathbf{m}_\sigma = (\mathbf{m}_{t_{k-1}},\mathbf{0}_{1 \times (Nd)})$} 
\STATE{Set $\mathbf{P}_\sigma = \begin{pmatrix}\mathbf{P}_{t_{k-1}} & \mathbf{0}_{n\times (Nd)} \\ \mathbf{0}_{(Nd) \times n} & I_{(Nd) \times (Nd)} \end{pmatrix}$}
\STATE{Generate $2(n+Nd)+1$ sigma points, with weighted mean $\mathbf{m}_\sigma$ and weighted covariance $\mathbf{P}_\sigma$}
\FOR{Each sigma point $\sigma^{(j)}$}
\STATE{Set $x_0 = \sigma^{(j)}_{1:n}$.}
\STATE{Set $\mathbf{Z}_{1:N} = \sigma^{(j)}_{n+1:n+(Nd)}$ (reshaping the right-hand side into a $d \times N$ matrix if appropriate).}
\STATE{Solve numerically Equation \eqref{randode}. Let $\bfx_T^{(j)}$ be the value of the solution after $T$ units of time.}
\STATE{Set $\mathcal{Y}_j = \mathbf{h}(\bfx_T^{(j)})$.}
\ENDFOR
\STATE{Predict the mean and variance of the incoming observation using \eqref{spmean} and \eqref{spvar}.}
\STATE{Upon arrival of the observation $\bfy_{t_k}$, update the mean $\mathbf{m}_{t_k}$ and variance $\mathbf{P}_{t_k}$ of the filtering distribution using \eqref{six}.}
\ENDFOR
\end{algorithmic}

\section{Numerical experiments}

A general analysis of the error induced by the series expansion approximation is difficult. One cannot easily exploit the usual tools from the theory of stochastic processes. In general, the truncated driving noise does not possess the Markov property, nor is it a martingale. The truncated driving noise is, however, a Gaussian process, and this structure is exploited in \cite{Friz2010} to demonstrate convergence to the true SDE. In the first part of this section we present a numerical investigation into the approximation error. 

We then compare the series expansion UKF with the cubature Kalman filter, which was found to be the most accurate and numerically stable amongst standard unscented transform-based filters in this context. There is already a considerable amount of theoretical and empirical evidence in the literature that sigma point methods outperform the extended Kalman filter, especially in tracking models such as the one described below (see, for example, \cite{Julier2000} \cite{Arasaratnam2010} \cite{Julier04unscentedfiltering}). In addition, one must compute the gradient of the drift function in order to implement the EKF. For some processes, this can be cumbersome. In contrast, our algorithm can be used as a `black box' filter. We compare our results with the UKF rather than the EKF to provide the most informative experiments. In these experiments, we use a Stratonovich-to-It\^{o} correction term to modify the dynamics of our approximation, so that the solution coincides with the It\^{o} dynamics \cite{Kloeden1999}.

\subsection{Approximation error}

In the general nonlinear case, analytic solutions for nonlinear multi-dimensional ordinary differential equations are rarely available in closed form. Hence, it is difficult to establish precise bounds on the error induced by the series expansion approximation. In this section we aim to investigate properties of the series expansion approximation numerically. In the example we consider, we will see that one can obtain a good approximation by truncating the series expansion \eqref{bmseries} after about ten terms.

We will test our approximation on a model of an aircraft turning in the $(x_1,x_3)$ plane. We model the motion of the aircraft using noisy dynamics that account for imperfections in the control system. The model also accounts for external forces such as wind that might affect the trajectory of the aircraft. We describe the state of the with a seven-dimensional vector $x_{1:7}$. The components $(x_{1},x_{3},x_{5})$ represent the position of the aircraft in rectangular cartesian coordinates, while the components $(x_2,x_4,x_6)$ describe its velocity. The number $x_7$ describes the rate at which the aircraft is turning in the $(x_1,x_3)$ plane.

The dynamics of the system are given by \eqref{SDE}, with
\beq
\mathbf{a}(x_{1:7}) =  
\begin{pmatrix}
x_2 \\
-x_7 x_4 \\
x_4 \\
x_7 x_2\\
x_6 \\
0 \\
0              
\end{pmatrix}
\eeq

\beq
\mathbf{b}(x_{1:7}) = \begin{pmatrix}
0 & 0 & 0 & 0 \\
\frac{\sqrt{1+ x_2^2}}{v} & \frac{\sqrt{1+ x_4^2}}{v_{xy}} & \frac{\sqrt{(1+ x_2^2)(1+ x_6^2)}}{v v_{xy}} & 0 \\
0 & 0 & 0 & 0 \\
\frac{\sqrt{1+ x_4^2}}{v} & -\frac{\sqrt{1+ x_2^2}}{v_{xy}} & \frac{\sqrt{(1+ x_4^2)(1+ x_6^2)}}{v v_{xy}} & 0 \\
0 & 0 & 0 & 0 \\
\frac{\sqrt{1+ x_6^2}}{v} & 0 & -\frac{v_{xy}}{v} & 0 \\
0 & 0 & 0 & 1
\end{pmatrix}
\eeq
Here, $v = \sqrt{1+ x_2^2 +  x_4^2 + x_6^2}$ and $v_{x y} = \sqrt{1+ x_2^2 + x_4^2}$. Nonlinearities arise from two sources in this system. Firstly, the state-dependent covariance matrix causes the system to deviate from Gaussianity. Second, the random evolution of the turn rate $\bfx_7(\cdot)$ causes the aircraft to behave erratically. As the variance of $\bfx_7(\cdot)$ grows, the system becomes more nonlinear and more non-Gaussian. A similar model was studied in \cite{Arasaratnam2010}, though in that case the diffusion matrix was assumed to be constant. Note that the state dependent covariance matrix makes It\^o-Taylor and Runge-Kutta discretisations difficult to implement.

\begin{figure}[h!]
\centering
\begin{subfigure}[]{0.3\textwidth}
\includegraphics[width=2.5in]{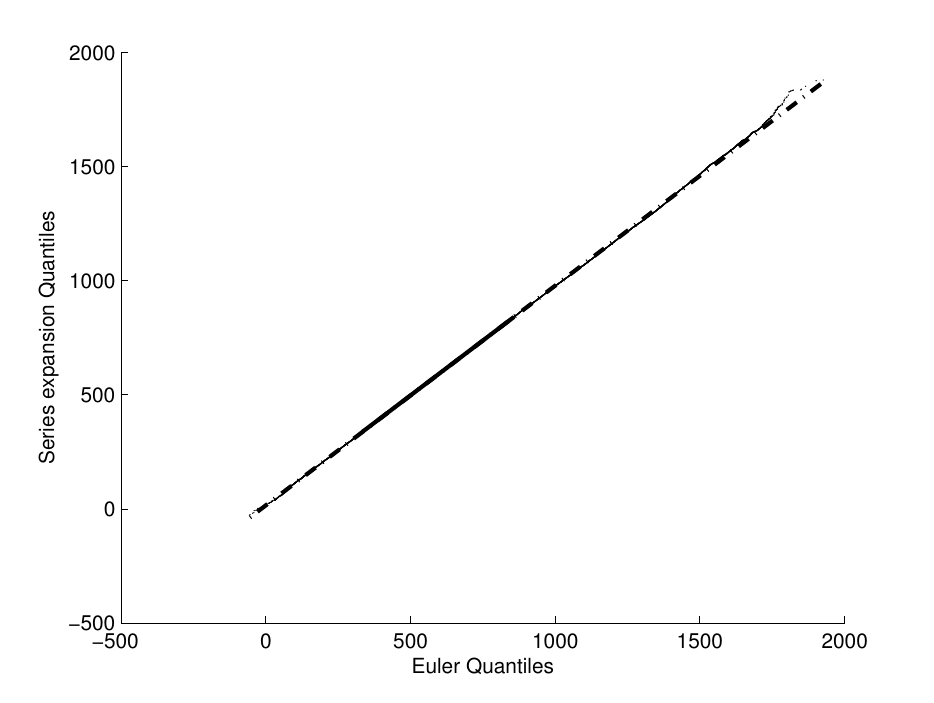}
\end{subfigure}
\begin{subfigure}[]{0.3\textwidth}
\includegraphics[width=2.5in]{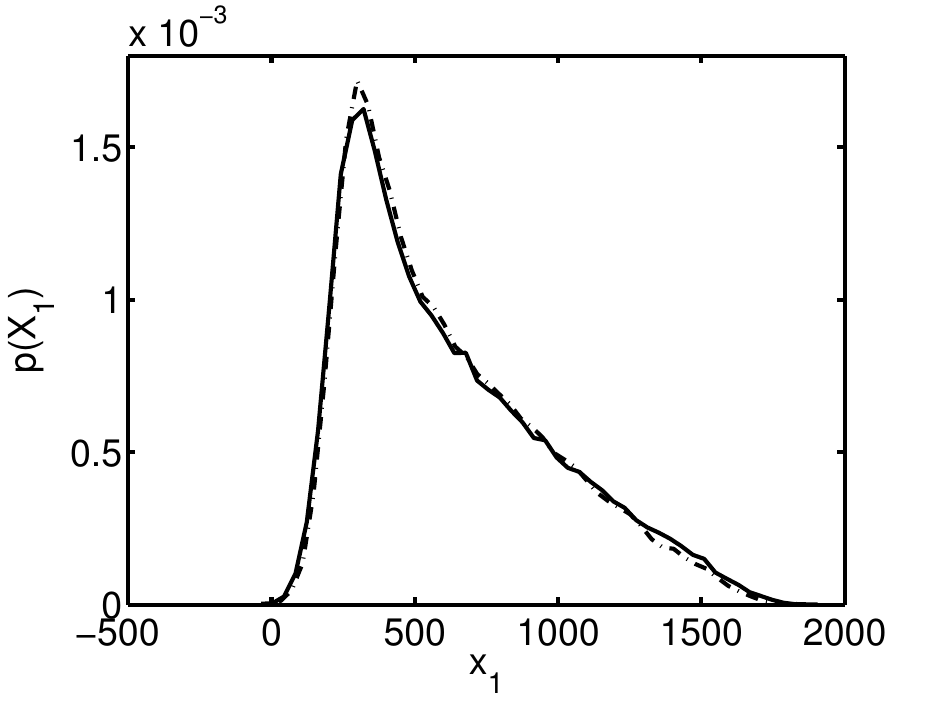}
\end{subfigure}
\caption{(Top) Q-Q plot of 100,000 samples from an Euler discretisation of $\bfx_1(T=8)$ versus 100,000 samples from the series expansion approximation. Linearity of the plot suggests the distributions are very similar. (Bottom) Density plots of the samples. Draws from the Euler scheme are plotted using the solid line, and draws from the series expansion scheme are represented by the broken line. We used the Fourier sine series as a basis, with $N = 10$.}
\label{fig_qq_density}
\end{figure}

In order to test the series expansion approximation, we simulated paths from $\bfx$ on the interval $[0,8]$. We set $\bfx_0 = (1000\text{ m},0\text{ m/s},2650\text{ m},150\text{ m/s},200\text{ m},0\text{ m/s},6\text{ deg/s})$, and $\text{Cov}(\bfw_t) = \text{Diag}(50,50,50,25)t$, resulting in a highly nonlinear process. We took $100,000$ simulations from the Euler-Maruyama scheme as ground truth, having set $\Delta t = .005$. The basis functions were defined by
\beq
\label{experimentbasis}
\phi_k(t) = \sqrt{\frac{2}{T}}\sin\left( \frac{(k - \frac{1}{2}) \pi t}{T}  \right),
\eeq
with $T=8$. We simulated $100,000$ paths from the series expansion approximation with $N = 1,4,6$ and $10$. The marginal means and standard deviations are shown in tables 1 and 2. Figure \ref{fig_qq_density} shows a Q-Q plot of the Euler simulation versus the series expansion simulation with $N=10$, together with a plot of both densities.  
\begin{table}
\begin{center}
    \begin{tabular}{ | l | c | c | c | c | c |}
    \hline
     & Euler & N = 1 & N = 4 & N = 6 & N = 10 \\ \hline
     $\ex[\bfx_1(t)]$ & 626 m & 549 m & 607 m & 612 m & 619 m \\ \hline
     $\ex[\bfx_2(t)]$ & -59 m/s & -91 m/s & -65 m/s & -63 m/s & -61 m/s \\ \hline
     $\ex[\bfx_3(t)]$ & 3588 m & 3689 m  & 3612 m & 3603 m &  3597 m \\ \hline
     $\ex[\bfx_4(t)]$ & 53 m/s & 82 m/s & 58 m/s & 56 m/s & 55 m/s \\ \hline
     $\ex[\bfx_5(t)]$ & 200 m & 200 m & 200 m & 200 m & 200 m  \\ \hline
     $\ex[\bfx_6(t)]$ & 0 m/s & 0 m/s & 0 m/s & 0 m/s & 0 m/s  \\ \hline
     $\ex[\bfx_7(t)]$ & 6 deg/s & 6 deg/s & 5.9 deg/s  & 6 deg/s & 5.9 deg/s \\ \hline
    \end{tabular}
\caption{Marginal mean values for $\bfx_{1:7}$ at $t = 8$ as computed by the Euler scheme and series expansion approximations}
\end{center}
\end{table}

\begin{table}
\begin{center}
    \begin{tabular}{ | l | c | c | c | c | c |}
    \hline
     & Euler & N = 1 & N = 4 & N = 6 & N = 10 \\ \hline
     $\text{Std}(\bfx_1(t))$ & 359 & 151 & 317 & 333 & 346 \\ \hline
     $\text{Std}(\bfx_2(t))$ & 90 & 61 & 86 & 88 & 89  \\ \hline
     $\text{Std}(\bfx_3(t))$ & 277 & 128  & 250 & 261 &  268 \\ \hline
     $\text{Std}(\bfx_4(t))$ & 93 & 66  & 90 & 91 & 92  \\ \hline
     $\text{Std}(\bfx_5(t))$ & 29 & 17 & 27 & 28 & 28  \\ \hline
     $\text{Std}(\bfx_6(t))$ & 6.4 & 5.7 & 6.2 & 6.3 & 6.3  \\ \hline
     $\text{Std}(\bfx_7(t))$ & 14.1 & 12.8 & 13.7  & 13.9  & 14.0 \\ \hline
    \end{tabular}
\caption{Marginal standard deviations for $\bfx_{1:7}(t=8)$ as computed by the Euler scheme and series expansion approximations}
\end{center}
\end{table}

\subsection{Filtering Experiments}
\label{filex}

As the nonlinearity of the system increases, the speed at which the filtering distribution deviates from Gaussianity should also increase. Intuitively, this means the amount of information that the conventional UKF `throws away' at each timestep grows with nonlinearity of the system. The series expansion method avoids this issue by targeting the predictive density at a given time directly without any intermediate projection onto the space of Gaussian distributions. As a result, we should expect the series expansion filter to outperform the conventional UKF in systems that are more highly nonlinear. 

To test this hypothesis, we set the covariance of the four-dimensional Brownian motion driving the aircraft model to $\text{Cov}(\bfw)(t) = \text{Diag}(10,0.2,0.2,Q_W^2)t$. The quantity $Q_W$ determines the variance of the turn rate of the aircraft. We use it as a proxy for the degree of nonlinearity of the system. We chose a number of values for  $Q_W$, ranging between $Q_W = 0.1$ and $Q_W = 1.1$. For each value of the variance, we simulated 1000 trajectories for the aircraft, running both filters on each trajectory. For each trajectory, the initial condition was drawn from a Gaussian distribution with mean $m_0 = (1000,0,2650,150,200,0,6)$. The standard deviation of each component was set to $100$, with the exception of the standard deviation of $\bfx_7(0)$ (recall that this notation denotes the seventh component of the vector at time 0, rather than the value at time 7). This was set to $0.1$. All components were assumed to be uncorrelated initially. 

For each trajectory, we simulated $n^{\text{obs}} = 20$ observations, spaced $T=8$ units of time apart. The observation function $\mathbf{h}$ models radar signals arriving at a dish. For this reason, we assume observations arrive in spherical coordinates, so that $\mathbf{h}$ is given by
\beq
\mathbf{h}(x_{1:7}) = 
\begin{pmatrix}
 \sqrt{x_1^2 + x_3^2 + x_5^2}\\
\tan^{-1}(x_3/x_1) \\
\tan^{-1}(x_5/\sqrt{x_1^2 + x_3^2}). 
\end{pmatrix}
\eeq
The covariance matrix of the observation noise was set to $\mathbf{R} = \text{diag}(50,0.1,0.1)$.

For the standard unscented Kalman filter, an It\^o-Taylor scheme such as the one proposed in  \cite{Arasaratnam2010} is impractical to implement as a result of the state-dependent noise. This is due to the presence of iterated stochastic integrals in which the integrand is a function of $\bfx_t$ (see \cite{Kloeden1999}). Even the simplified order 2.0 It\^o-Taylor scheme proposed in \cite{Kloeden1999} is cumbersome to implement. For an $n$-dimensional process, we need to compute $n^2 + 2n + 1$ terms involving derivatives of the coefficient functions (in our case, $n=7$ so this means 64 terms). The simplified scheme also involves a number of Bernoulli random variables, and it is not immediately clear how one would incorporate these into an unscented filter.

We chose to use the limiting scheme first proposed in \cite{Sarkka2007}. The system of ODEs \eqref{predictivemomentsode} was solved by a fourth order Runge-Kutta scheme. The number of Runge-Kutta steps used did not appear to affect the error appreciably. However, with a large step size the predicted covariance can fail to be positive definite, which causes the filter to diverge. We found that a good compromise between computational cost and the divergence issue was to choose a smaller step-size for more highly nonlinear parameter settings. For this reason, we used $200 Q_\bfw$ steps per unit time.

The system of ordinary differential equations \eqref{randode} defining the series expansion method was solved numerically using the Dormand-Prince Runge-Kutta method. This is the default ODE solver implemented in MATLAB. It is an adaptive algorithm, and the number of timesteps used depends on the integrand.

Run-time of either algorithm depends on a number of factors. The main factors that determine computation time are the numerical method used (and the number of timesteps in that method), and the number $N$ of series expansion terms. In our setup, we found that the series expansion method could run anywhere from twice as fast to four times slower than the standard unscented filter. We stress, however, that no effort was made to push either method to the limit of efficiency.

For the standard unscented filter, we set $\alpha = 1$, $\kappa = 0$ and $\beta = 0$. This choice of tuning parameters is also known as the cubature Kalman filter \cite{Arasaratnam2009, Arasaratnam2010}. Various other parameter settings produced similar results, though these settings were most stable and most accurate.

For the series expansion method, we used the orthonormal basis \eqref{experimentbasis} with $T = 8$, and used $N=8$ basis functions for each component of the Brownian motion. 

The series expansion filter takes one large step instead of many small ones. As such, one can expect that the target distribution is less like a Gaussian distribution. We found that `tweaking' the standard parameters slightly improved performance, though not dramatically. We set $\alpha = 1, \kappa = -32, \beta = 0$ so that $\lambda = 7$. Our motivation for this choice is given in Section \ref{discussion}.

For any given sample path, we compute the root mean squared error for the position, velocity and turn rate:
\beq
\label{rmse}
\epsilon_{\text{c}} = \sqrt{\frac{1}{n^{\text{obs}} l}\sum_{k=1}^{n^{\text{obs}}}  \left( \bfx_c (t_k)  - \mathbf{m}_c(t_k) \right)^\top \left( \bfx_c (t_k)  - \mathbf{m}_c(t_k) \right) },
\eeq
This results in a collection $\{\mathbf{\epsilon}^{(i)}\}_{i \leq n^{\text{obs}}}$ of vectors recording the errors for each sample path. Here, $\mathbf{m}_c(t_k)$ is the mean of the filtering distribution at time $t_k$. The value of $c$ depends on the error component. For position errors, $c = (1,3,5)$. For velocity errors, $c = (2,4,6)$, and for turn rate errors, $c = 7$. We set $l = 3$ for the position and velocity errors and $l = 1$ for the turn rate error. Mean filter errors and divergences are reported in Table \ref{errtable}. A filter was deemed to have diverged if the RMSE position error was greater than 1 km. When this occurred, the corresponding value of $\epsilon^{(i)}$ was not included in the average.

\begin{figure}
\centering
\begin{subfigure}[]{0.3\textwidth}
\includegraphics[width=2.5in]{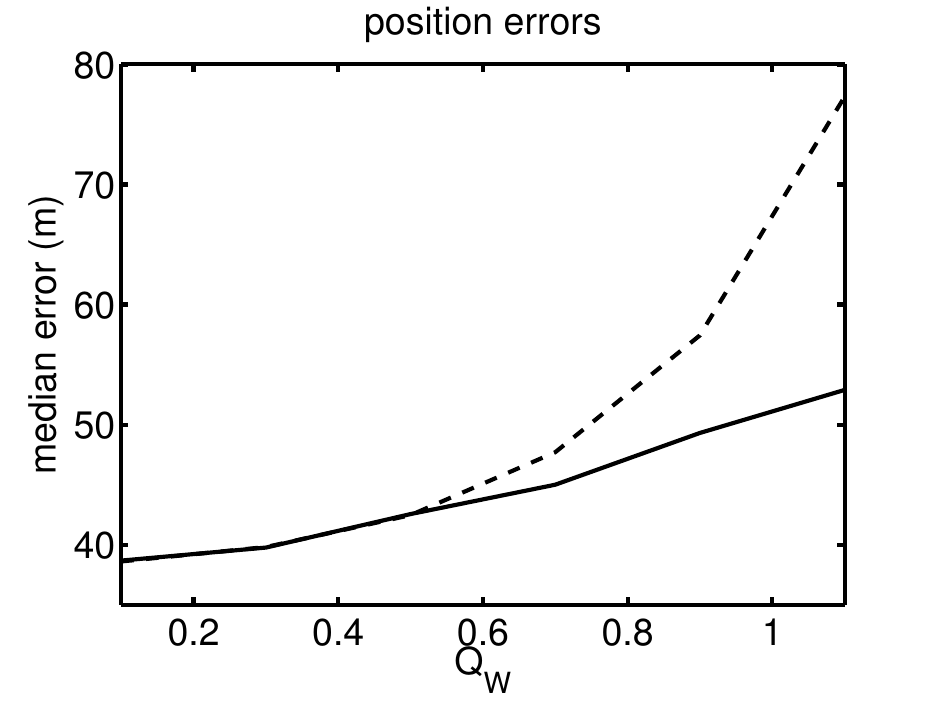}
\end{subfigure}
\begin{subfigure}[]{0.3\textwidth}
\includegraphics[width=2.5in]{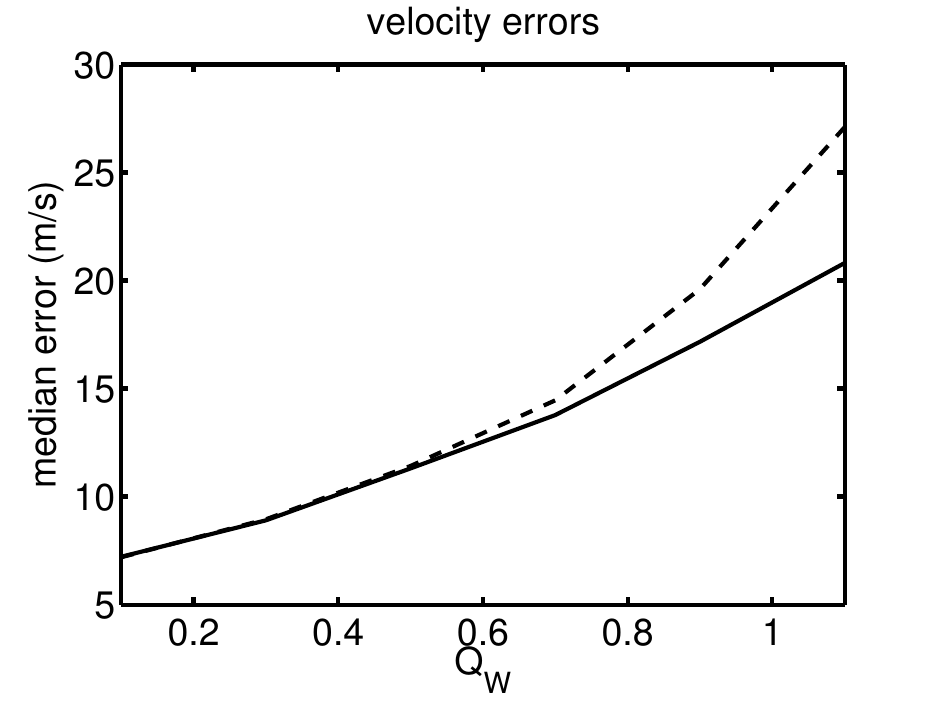}
\end{subfigure}
\begin{subfigure}[]{0.3\textwidth}
\includegraphics[width=2.5in]{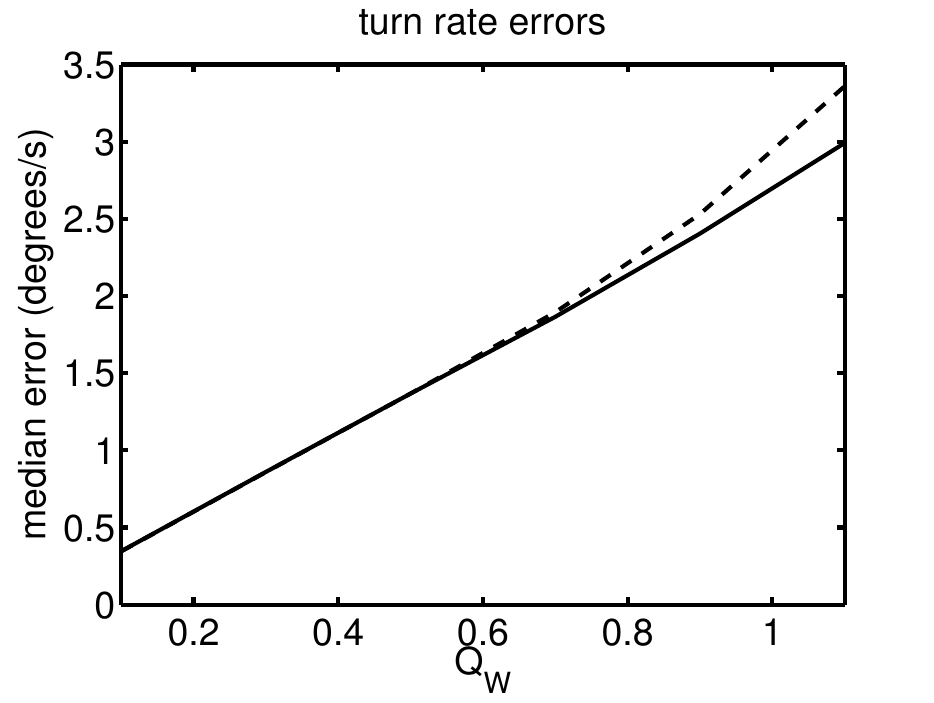}
\end{subfigure}
\caption{The $x$-axis shows the diffusion coefficient $Q_W$ of the Brownian motion driving $\bfx_7(t)$. We use this as a measure of the nonlinearity of the system. For a range of values of $Q_W$, we simulated 1000 trajectories of the signal, observed with noise. We plot median values of the error for the unscented Kalman filter (dotted line) and series expansion filter (solid line).}
\label{errmedian}
\end{figure}

\begin{figure}
\centering
\begin{subfigure}[]{0.3\textwidth}
\includegraphics[width=2.5in]{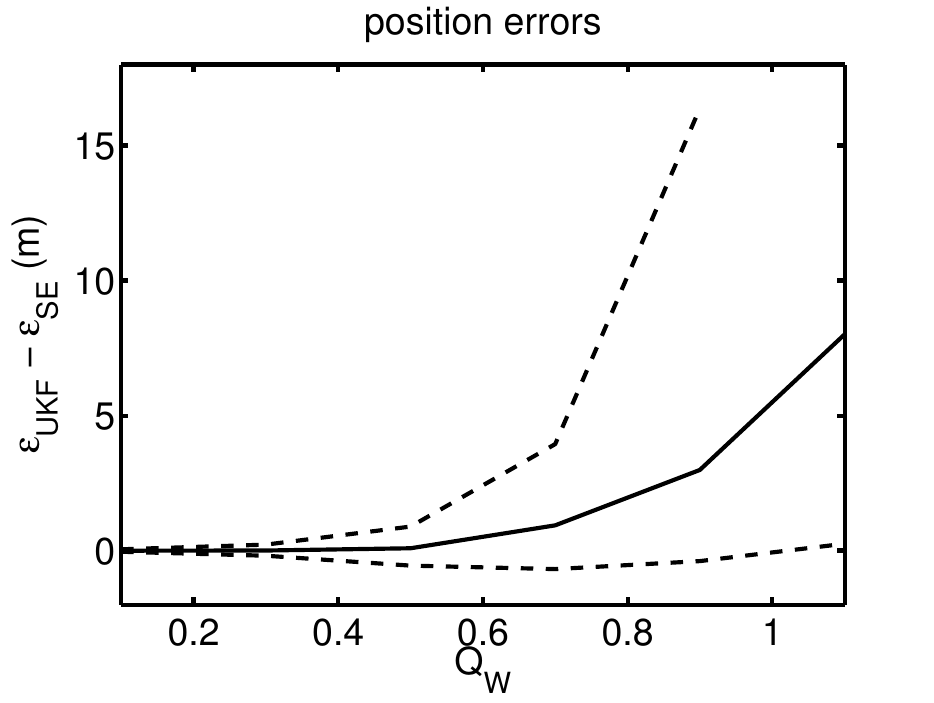}
\end{subfigure}
\begin{subfigure}[]{0.3\textwidth}
\includegraphics[width=2.5in]{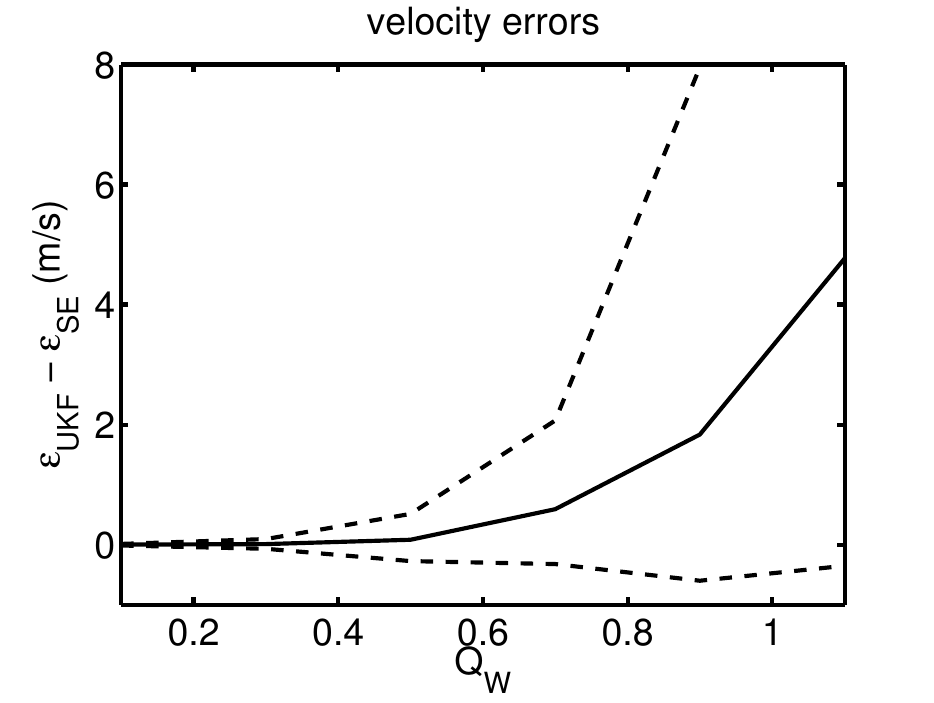}
\end{subfigure}
\begin{subfigure}[]{0.3\textwidth}
\includegraphics[width=2.5in]{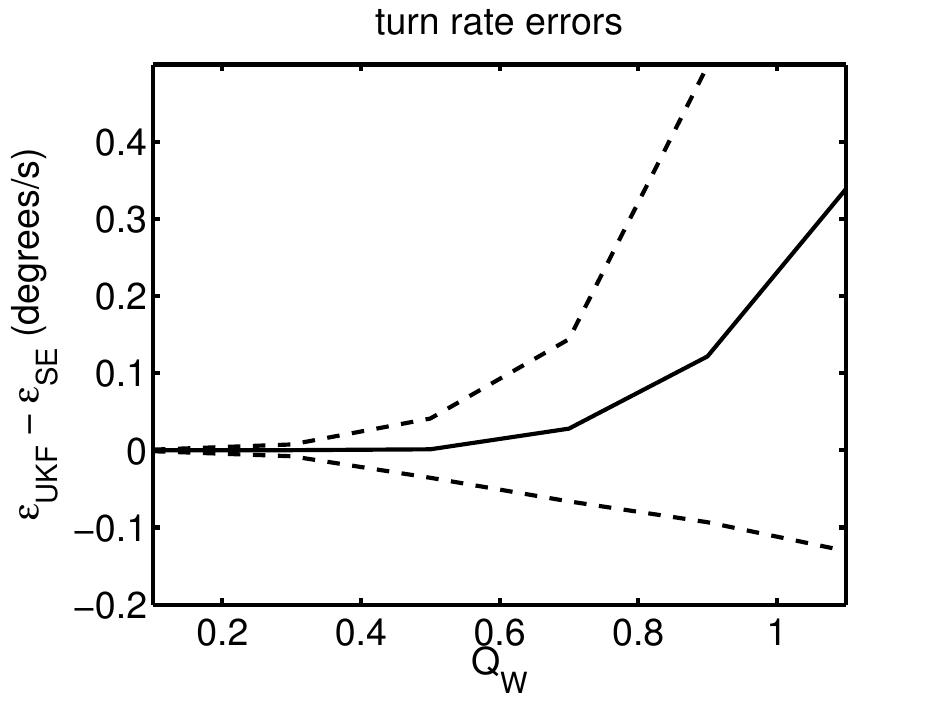}
\end{subfigure}
\caption{The $x$-axis shows the diffusion coefficient $Q_W$ of the Brownian motion driving $\bfx_7(t)$. We use this as a measure of the nonlinearity of the system. For a range of values of $Q_W$, we simulated 1000 trajectories of the signal, observed with noise. We plot median values of the difference in error between the unscented Kalman filter and series expansion Kalman filter (solid line), together with the first and third quartiles (dashed lines). Errors were computed seperately for position, velocity and turn rate of the aircraft. The last point in the upper range is omitted because its inclusion would skew the scaling in the image. Values for these points can be found in Section \ref{filex}.}
\label{errdiff}
\end{figure}

Both the series expansion filter and unscented filter can diverge and lose track of the signal, in which case the error becomes very large. Even if divergences are discarded, a few large errors can still dominate the average. For this reason, we report the median over all runs of the absolute error for each component in Figure \ref{errmedian}. We report quartiles of the empirical distribution of $\epsilon_{\text{UKF}}^{(i)} - \epsilon_{\text{SE}}^{(i)}$ in Figure \ref{errdiff}. 

Figure \ref{errmedian} shows the median values of the difference in errors together with the first and third interquartiles. The third interquartile corresponding to $Q_W = 1.1$ is excluded because the plot could not be scaled appropriately. For the position, the value is $77m$ . For the velocity, $ 67 \text{m/s}$, and for the turn rate, $7.8$ degrees/s.

\begin{table}[h!]
\begin{center}
    \begin{tabular}{ | l | c | c | c |}
    \hline
     $Q_W$ & .1 & .3 & .5 \\ \hline
     RMSE UKF (divs)& 49.9 m (1) & 49.7 m (3) & 55.0 m (12) \\ \hline
     RMSE SE-UKF (divs)& 49.9 m (1) & 49.8 m (4) & 56.4 m (7) \\ \hline
      \multicolumn{1}{c}{}
      \\ \hline
     $Q_W$  & .7 & .9 & 1.1 \\ \hline
     RMSE UKF (divs)&  66.9 m (28) & 92.2 m (75) & 136.7 m (107)\\ \hline
     RMSE SE-UKF (divs)& 63.4 m (17) & 71.5 m (20) & 83.5 m (50)\\ \hline
     
    \end{tabular}
\caption{Mean position errors and divergences for 1000 runs of the filter. Larger values of $Q_W$ result in more erratic trajectories. The filter was deemed to have diverged if the position error was greater than 1km, or if the filter failed due to the appearance of a non-positive definite covariance matrix. Divergent runs were not included in the average. The number of divergences is reported in parentheses}
\label{errtable}
\end{center}
\end{table}

Choice of basis functions made minimal difference in this experiment. We re-ran the experiment using $N=8$ Haar wavelet functions instead of sinusoidal basis functions. Results for the most nonlinear setting $Q_W = 1.1$ are shown in Table \ref{basistable}. Filtering errors for both sets of basis functions were close to one another. This is because the Gaussian approximation and tuning parameters of the unscented transform have a larger effect on the filter than specifics of the series expansion approximation.

\begin{table}
\begin{center}
    \begin{tabular}{ | l | c | c | c | c | c |}
    \hline
     Basis & Pos. Error & Vel. Error & Turn Error\\ \hline
     Sine & 53.4 m & 20.8 m/s & 300 deg/s \\ \hline
     Haar & 53.6 m & 20.9 m/s & 301 deg/s \\ \hline
    \end{tabular}
\caption{Error induced by using a Haar wavelet basis versus error from a sinusoidal basis. Median error from 1000 runs of the filter. We used the most highly nonlinear setting, $Q_W = 1.1$.}
\label{basistable}
\end{center}
\end{table}

Surprisingly, we found that choosing the symmetric square root of $P_t$ (that is, the matrix that satisfies $\mathbf{S}^2 = \mathbf{P}_t$, implemented in MATLAB as sqrtm$()$) instead of the Cholesky decomposition improved the accuracy of our algorithm considerably (though this choice did not improve performance of the standard UKF). The choice of matrix square root is known to affect fourth-order and higher terms in the Taylor expansion of the transition function $\mathbf{f}$ \cite{Julier04unscentedfiltering}. This is in agreement with our intuition: the transition function in the UKF is locally linear, and hence can be approximated with a low-order Taylor series. On the other hand, the series expansion filter uses a more nonlinear transition function and one must consider higher order terms.

\subsection{Series expansion step size}
\label{stepsize}
In the prediction step of the standard unscented filter, one discretises the process $\bfx$, and iteratively applies the unscented transform at each timestep. The aim is to estimate the mean and covariance of $\bfx_t$ at some time $t$, given an appropriate initial condition. Repeated applications of the unscented transform at each timestep induce error in this estimate. We will refer to error of this nature as `projection error'.

On the other hand, the error in the SE-UKF comes from the error induced by the series expansion approximation, coupled with the error induced by a single application of the unscented transform. Error also accrues from numerical solution of the ODE, but in our experiment, this is negligible compared to other sources of error. Empirically, we observe that the accuracy of the series expansion approach improves with the number $N$ of basis functions that we use, and deteriorates with the time $T$ between observations. We will refer to error induced by the series expansion as `approximation error'. 

In one sense, these two approaches represent two extremes of a more general framework. For example, we might use the series expansion approximation to estimate the mean and variance of $\bfx_{T/2}$. We could then form a Gaussian approximation of its distribution, and use this as the initial condition (starting at time $T/2$) for a second application of the series expansion trick to estimate the mean and covariance of $\bfx_T$. In effect, we reduce the approximation error at the cost of increasing the projection error.

In order to investigate the effect of trading approximation error for projection error, we ran the filtering experiment of Section \ref{filex} using the most nonlinear setting, $Q_W = 1.1$. Recall that the time interval between observations was $T=8$ seconds. We divided this interval into $K$ subintervals of length $T/K$. At the end of each subinterval, we re-initialised the series expansion approximation, using as initial condition the mean and variance computed at the previous sub-interval.

\begin{table}
\begin{center}
    \begin{tabular}{ | l | c | c | c | c | c |}
    \hline
     K & Pos. Error & Vel. Error & Turn Error\\ \hline
     1 & 54.0 m & 21.1 m/s & 293 deg/s \\ \hline
     2 & 51.6 m & 20.2 m/s & 291 deg/s   \\ \hline
     4 & 51.6 m & 20.6 m/s & 290 deg/s \\ \hline
     8 & 55.5 m & 21.6 m/s & 294 deg/s  \\ \hline
     16 & 61.7 m & 22.8 m/s & 304 deg/s  \\ \hline
     32 & 69.3 m & 24.2 m/s & 322 deg/s  \\ \hline
    \end{tabular}
\caption{The effect of trading approximation error for projection error. Median errors over 1000 runs of the filter. Rows are indexed by number $K$ of projections per observation. We used the most challenging parametrisation $Q_W = 1.1$ to generate the data. Observe that results for $K = 32$ correspond closely to the errors for the standard UKF in Figure \ref{errmedian}.}
\label{projtable}
\end{center}
\end{table}

Table \ref{projtable} shows that one can reduce the error slightly by repeatedly employing the series expansion approximation over a shorter timescale, thus trading approximation error for projection error. As the number $K$ of projections becomes large, the error grows to match that of the standard UKF.

\section{Discussion and conclusions}
\label{discussion}
In this paper, we have presented a Gaussian filter based on the series expansion approximation. The novel contributions of this paper focus on improving the predictive distribution, so it is straightforward to construct a smoother using similar methods: for example one can use the unscented smoother \cite{Sarkka2008} or Gaussian smoother \cite{Sarkka2010} directly.

Two questions follow naturally from this work. Firstly, how does one choose parameters for the unscented transform in a sensible way? Secondly, what basis functions should one use in the series expansion? In most cases the optimal solution for either question is likely to be very difficult to compute. 

All filters based around the unscented transform must somehow deal with the first issue. Various heuristics can be found in the literature on how one might choose the tuning parameters: see, for example \cite{Turner2012}, \cite{Gustafsson2012}. In some cases, a poor choice of tuning parameters can cause the covariance matrix in the prediction step to fail to be positive definite. This causes the filter to diverge. 

When using a common set of tuning parameters ($\alpha = 1, \kappa = 3-n, \beta = 2$, where $n$ is the dimensionality of the system \cite{Wan2000}), we found the matrix degeneracy problem to occur in both the series expansion filter (about 1\% of runs) and the standard unscented filter (about 10\% of runs). This is a known issue when using these settings in a high-dimensional context \cite{Wan2000}. We found that increasing $\kappa$ slightly to $\kappa = 5-n$ in the series expansion filter removed the divergence issue without affecting performance. On the other hand, the cubature Kalman filter settings ($\alpha = 1, \kappa = 0, \beta = 0$) performed poorly for the series expansion filter. This is because the higher dimensionality of the system causes the sigma points to be spread far out from the mode (that is, $\lambda = \alpha^2(n + \kappa) - n$ is large).

We also compared our algorithm to the third-order Gauss-Hermite
Kalman filter (GHKF). This algorithm also exhibited numerical instability,
with the predicted covariance matrix often failing to be positive definite.
When we discarded test runs on which the GHKF diverged, we found that
our algorithm performed comparably to the GHKF. This is despite the fact that the cost of the GHKF scales exponentially with dimension. In the present setting, the GHKF used $3^7 = 2187$ sigma points, and required several days of computation time to perform a comparison for a single value of $Q_W$.

We now address the issue of the choice of orthonormal basis. We performed the same filtering experiments using a sinusoidal basis, and a basis of Haar wavelets. Results were similar in both cases. Our explanation for this is that we already induce significant error by assuming the filtering distribution is Gaussian. This error is significantly larger than the error induced by the series expansion approximation, so the latter error is difficult to detect. 

In any case, we recommend choosing a basis that converges uniformly to Brownian motion, so that one has concrete theoretical guarantees of convergence without the need to invoke rough path theory.

For completeness, we outline one more possible strategy for choosing basis functions for the series expansion. When the SDE is linear, it is possible to construct a set of basis functions such that the series expansion approximation is error-free. In the univariate case, when $b$ is constant the solution to \eqref{randode} is given by

\beq
\hat{X}_T = X_0 \exp(a T) + \sum_{i=1}^N Z_i \int_0^T \exp(a (t-u))b \phi_i(u)du.
\eeq

If we set $\phi_1(u) = \exp(-a u)/\|\exp(-a \cdot)\|$, then the integral on the right disappears for all $i>1$. This is because all other basis functions are orthogonal to $\exp(-a \cdot)$ by construction. Thus the approximation is exact at time $T$ (though not necessarily at earlier times $t<T$). The argument in the multivariate case is similar, though slightly more complicated.

To choose a set of basis functions in a non-linear setting, one can first construct a linear approximation to the non-linear problem. One then computes the optimal basis functions for the linearized dynamics as above. However, in out numerical tests we found that these basis functions were prone to numerical instability, and furthermore they do not come with a guarantee of uniform convergence. We note that this may be a useful strategy in filtering problems that are `almost' linear.

\appendix[Convergence of the approximation]
\label{sdeconverge}
We now discuss asymptotic convegence of the approximation \eqref{randode} to the solution of the true SDE. Wong and Zakai \cite{Wong1965} showed that in the univariate case, under mild technical assumptions, the solution of \eqref{randode} converges to the Stratonovich solution of the SDE that is being approximated.

We use the circle notation to denote Stratonovich integration. Recall that a Stratonovich SDE
\beq
d\bfx_t = \bfa(\bfx_t)dt + \bfb(\bfx_t)\circ d\bfw_t
\eeq
can be converted to an It\^o SDE and vice versa using the relationship
\beq
\int_0^t \bfb(\bfx_t)\circ d\bfw_t = \int_0^t \bfb(\bfx_t)d\bfw_t + \int_0^t \mathbf{c}(\bfx_t) dt,
\eeq
where the integral on the left is in the Stratonovich sense, and the $i$-th component of the vector $\mathbf{c}$ satisfies
\beq
\mathbf{c}^i(x) = -\frac{1}{2} \sum_{j=1}^n \sum_{k=1}^d \bfb^{j,k}(x) \frac{\partial \bfb^{i,k}}{\partial x_j}(x).
\eeq
In other words, the Stratonovich solution of an SDE is equivalent to the It\^o solution with a modified drift.

The issue of convergence in the multidimensional setting is somewhat more involved than in the univariate case. In general, if  $\{\bfw_{n}\}$ is a sequence of piecewise smooth processes converging to a Brownian motion, one cannot guarantee  $\{\bfw_{n}\} \rightarrow \bfw$ implies that the sequence of approximate differential equations converges to the Stratonovich solution of the SDE. One must impose some extra conditions on the so-called `Levy area' of the Brownian approximations. Let $W^j_{n,u}$ be the $j$-th component of $\bfw_n$ at time $u$, and let $W^j_u$ be the $j$-th component of $\bfw$ at time $u$. Define a set of processes

\beq
S^{ij}_{n,t} = \int_0^t \left(W^j_u - W^j_{n,u}\right)dW^i_{n,u} - \delta_{ij}t.
\eeq 
Many results about the convergence issue are known in the mathematical literature. For example, suppose the following conditions hold with probability 1 for all $\kappa$ less than some positive number $\gamma$:
\beq
\label{uc}
\sup_{u \leq T} \| \bfw_u -\bfw_{n,u} \| = O(n^{-\kappa}), 
\eeq
\beq
    \sup_{u \leq T} \| S^{ij}_{n,u} \| = O(n^{-\kappa}),
\eeq
\beq
    \int_0^T \left \| \frac{d}{du} S^{ij}_{n,u} \right \| du = O(\log^{\delta}(n)) \qquad \forall \delta > 0.
\eeq

The thesis of Schmatkov \cite{Schmatkov2005} showed that under these assumptions,

\beq
\sup_{u \leq T} \| \bfx_u - \hat{\bfx}_u \| = O(n^{-\kappa}). 
\eeq
See \cite{Gyongy2006} for an analogous result about stochastic partial differential equations.

In general, there is no guarantee that the sequence of partial sums in \eqref{dbmseries} converges uniformly (so \eqref{uc} is not necessarily satisfied). If one chooses the Haar wavelets as an orthonormal basis in which to expand the driving Brownian motion, then convergence is indeed uniform: in fact this choice corresponds to the L\'{e}vy-Ciesielski construction of Brownian motion \cite{McKean1969}. For a general choice of basis functions, one can show that $\hat{\bfx} \rightarrow \bfx$ provided that the processes are interpreted as \emph{rough paths} \cite{Friz2010}.

% or


\section*{Acknowledgment}

Simon Lyons was supported by Microsoft Research, Cambridge.




\bibliographystyle{plain}
%\bibliography{IEEEabrv,sde.bib}
\bibliography{sde}


%\newpage



%\vfill


\end{document}